\documentclass[prl,showpacs,superscriptaddress,twocolumn,amssymb,amsfonts]{revtex4}
\usepackage{amsmath}
\usepackage{bm}
\usepackage{epsfig}
\usepackage{graphics}

\def\beq{\begin{equation}}
\def\eeq{\end{equation}}
\def\beqn{\begin{eqnarray}}
\def\eeqn{\end{eqnarray}}

\def\E{{\bf E}}
\def\B{{\bf B}}

\def\M{{\bf M}}

\def\del{\partial}
\def\F{{\cal F}}
\def\A{{\cal A}}
\def\ket#1{\vert #1 \rangle}
\def\ev#1{\langle #1 \rangle}

\def\me#1#2#3{\langle #1 \vert #2 \vert #3 \rangle}
\def\inh{^{\rm (in)}}
\def\intbz{\int_{\rm BZ}\frac{d^3k}{(2\pi)^3}}

\begin{document}
\title{Magnetoelectric polarizability and axion electrodynamics in crystalline insulators}
\author{Andrew M. Essin}
\affiliation{Department of Physics, University of California,
Berkeley, CA 94720}
\author{Joel~E.~Moore}
\affiliation{Department of Physics, University of California,
Berkeley, CA 94720} \affiliation{Materials Sciences Division,
Lawrence Berkeley National Laboratory, Berkeley, CA 94720}
\author{David Vanderbilt}
\affiliation{Department of Physics and Astronomy, Rutgers University, Piscataway, NJ 08854}
\date{\today}

%%%%%%%%%%%%%%%%%%%%%%%%%%%%%%%%%%%
\marginparwidth 2.7in
\marginparsep 0.5in
% \def\dvm#1{\marginpar{\small DV: #1}}
% \def\jem#1{\marginpar{\small JM: #1}}
% \def\aem#1{\marginpar{\small AE: #1}}
%uncomment next lines to have no commentaries
% \def\dvm#1{}
% \def\jjm#1{}
% \def\aem#1{}
%
\def\scr{\scriptsize}
%%%%%%%%%%%%%%%%%%%%%%%%%%%%%%%%%%%

\begin{abstract}
The orbital motion of electrons in a three-dimensional solid can
generate a pseudoscalar magnetoelectric coupling $\theta$, a 
fact we derive for the single-particle case using a recent theory 
of polarization in weakly inhomogeneous materials.  This 
polarizability $\theta$ is the same parameter that appears in the 
``axion electrodynamics'' Lagrangian 
$\Delta{\cal L}_{EM} = (\theta e^2 / 2 \pi h) {\bf E} \cdot {\bf B}$,
which is known to describe the unusual magnetoelectric properties of the 
three-dimensional topological insulator ($\theta=\pi$).  We compute 
$\theta$ for a simple model that accesses the topological insulator 
and discuss its connection to the surface Hall conductivity.
The orbital magnetoelectric polarizability can be generalized to 
the many-particle wavefunction and defines the 3D topological 
insulator, like the IQHE, in terms of a topological ground-state 
response function.
\end{abstract}


\maketitle

Magnetoelectric couplings in solids have recently been the subject of intense
experimental and theoretical
investigations~\cite{spaldin-s05,fiebig-jpdp05,expttheta}.
A quantity of central importance is the linear magnetoelectric
polarizability $\alpha_{ij}$ defined via
\beq
\alpha_{ij}=
\frac{\partial M_j}{\partial E_i}\Big|_{{\bf B}=0} =
\frac{\partial P_i}{\partial B_j}\Big|_{{\bf E}=0}
\label{eq:alpha}
\eeq
where $E$ and $B$ are electric and magnetic fields, $P$ and $M$ are the
polarization and magnetization, and the equality can be obtained from commuting derivatives of an appropriate free energy.
In general the tensor $\alpha$
has nine independent components, and can be decomposed as
\beq
\alpha_{ij}=
\tilde{\alpha}_{ij}
+
\frac{\theta e^2}{2\pi h} \, \delta_{ij}
\label{eq:alpha-decomp}
\eeq
where the first term is traceless and the second term,
written here in terms of the dimensionless parameter $\theta$,
is the pseudoscalar part of the coupling.
Here we focus on magnetoelectric coupling resulting from the orbital (frozen-lattice) magnetization and polarization, which we label the orbital magnetoelectic polarizability (OMP).

In field theory, the pseudoscalar OMP coupling
is said to generate ``axion electrodynamics'' \cite{wilczekaxion},
and corresponds to a Lagrangian of the form ($c=1$)
\beq
\Delta {\cal L}_{EM} = \frac{\theta e^2}{2\pi h} {\bf E} \cdot {\bf B} =
\frac{\theta e^2}{16\pi h} \epsilon^{\alpha \beta \gamma \delta}
	F_{\alpha \beta} F_{\gamma \delta}.
\label{axioncoupling}
\eeq
An essential feature of the axion theory is that, when the axion
field $\theta(\bm{r},t)$ is constant, it plays no role
in electrodynamics; this follows because $\theta$ couples
to a total derivative, $\epsilon^{\alpha \beta \gamma \delta}
F_{\alpha \beta} F_{\gamma \delta} = 2 \epsilon^{\alpha \beta
\gamma \delta} \partial_\alpha (A_\beta F_{\gamma \delta})$, and
so does not modify the equations of motion.  However, the presence
of the axion field can have profound consequences at surfaces and
interfaces, where gradients in $\theta(\bm{r})$ appear.

A second essential feature is that electrodynamics
is invariant under
$\theta \rightarrow \theta+2 \pi$~\cite{wilczekaxion}.  In order to reconcile this
peculiar fact with the phenomenology of the magnetoelectric effect,
observe that the axion coupling can alternatively be described
in terms of a \emph{surface Hall conductivity} $\sigma_\mathrm{H}$
whose value $\theta e^2/2\pi h$ is determined by bulk properties,
but only modulo the quantum $e^2/h$.  More generally, at an interface
between two samples,
% (one of which could be free space),
$\sigma_\mathrm{H}
= (\theta_1-\theta_2+2\pi r)e^2/2\pi h$, where the integer $r$ depends on
the details of the interface.
Recall that, in general, a 2D gapped crystal
has an integer TKNN invariant $C$ in terms of which the
its Hall conductivity is $\sigma_\mathrm{H}=Ce^2/h$~\cite{tknn}.
The ``modulo $e^2/h$'', or integer $r$, discussed above corresponds
to modifying the surface or interface by adsorbing a surface layer of nonzero $C$.

When time-reversal ($T$) invariance
is present, the TKNN invariants vanish, but other invariants
arise that have been the focus of much recent work.
In 2D there is a $\mathbb{Z}_2$ invariant~\cite{km2} distinguishing
``ordinary'' from ``$\mathbb{Z}_2$-odd'' insulators, with
``quantum spin Hall'' states~\cite{kane&mele-2005,zhangscience1}
providing examples of the latter.  In 3D there is a similar
invariant~\cite{moore&balents-2006,rroy3d,fu&kane&mele-2007}
that can be computed either from the 2D invariant on certain
planes~\cite{moore&balents-2006} or from an index involving the eight $T$-invariant
momenta~\cite{fu&kane&mele-2007}.  If this is odd, the
material is a ``strong topological insulator'' (STI).
In the context of the OMP, note that $T$
maps $\theta\rightarrow-\theta$; the ambiguity of $\theta$ modulo $2\pi$
then implies that $T$ invariance is consistent
with either $\theta=0$ or $\theta=\pi$, with the latter corresponding
to the STI~\cite{qilong}.  Note
that if $T$-invariance extends to the surfaces, these become
metallic by virtue of topologically protected edge
states,
as observed experimentally for the
Bi$_{0.9}$Sb$_{0.1}$ system~\cite{hsieh}.  If the surface
is gapped by a $T$-breaking perturbation,
then $\sigma_\mathrm{H}=e^2/2h$ modulo $e^2/h$ at the surface of a 
STI~\cite{wilczekaxion,fu&kane2-2006,qilong}.

In the noninteracting case, a Berry-phase expression for $\theta$
has been given in terms of the bulk bandstructure by Qi, Hughes,
and Zhang~\cite{qilong} by integrating out electrons in one
higher dimension.  Defining the Berry connection
$\A^{\mu\nu}_j=i\me{u_\mu}{\del_j}{u_\nu}$ where $\ket{u_\nu}$ is the
cell-periodic Bloch function of occupied band $\nu$ and
$\del_j=\partial/\partial k_j$, they obtain
\beq
\theta=\frac{1}{2\pi}\int_{\rm BZ} d^3k \; \epsilon_{ijk}\,
{\rm Tr}[\A_i\del_j\A_k-i\frac{2}{3}\A_i\A_j\A_k]
\label{theta}
\eeq
where the trace is over occupied bands.
Note that wavevector-dependent unitary transformations 
(``gauge transformations'') on the set of occupied wave 
fuctions cannot affect bulk physical properties.

In the present letter, we first provide an alternate derivation
of Eq.~(\ref{theta}) for the OMP.  Our derivation clarifies
that $\theta$ is a polarizability and in fact describes a contribution
to magnetoelectric polarizability from extended orbitals.
The derivation
follows from an extension \cite{xiao} of the
Berry-phase theory of polarization~\cite{ksv} to the case of
slow spatial variations of the Hamiltonian.  (Indeed, the OMP angle
$\theta$ is a bulk property in exactly the same sense as
electric polarization~\cite{ortizmartin,ksv}.)  We find that the OMP can be generalized
to the interacting case and calculated from the many-particle
wavefunction, even though Eq.~(\ref{theta}) is not valid;
this reflects a subtle difference between OMP and polarization.
Explicit numerical calculations on
model crystals are presented to validate the theory, establish
the equivalence of Eq.~(\ref{theta}) to the prior definition, and
illustrate how a non-zero $\theta$ corresponds to a
``fractional'' quantum Hall effect at the surface of a magnetoelectric
or topological insulator~\cite{wilczekaxion,fu&kane2-2006,qilong}.

From Eq.~(\ref{eq:alpha}) it is evident that
the OMP can be viewed in several ways.
(i) It describes the {\it electric} polarization
arising from the application of a small {\it magnetic} field.
(ii) It describes the orbital {\it magnetization}
arising from the application of a small {\it electric} field.
(iii) It also gives the (dissipationless) {\it surface Hall conductivity}
$\sigma_\mathrm{H}$
at the surface of the crystal, provided that the surface
is insulating. 
Note that (iii) follows from (ii): for a surface
with unit normal $\hat{\bf n}$ and electric field $\E$, the
resulting surface current ${\bf K}=\M\times\hat{\bf n}$ is proportional
to $\E\times\hat{\bf n}$.
There is an elegant analogy here to the case of electric polarization,
where the surface charge of an insulating surface is determined,
modulo the quantum $e/S$, by the bulk bandstructure alone
($S$ is the surface cell area).

The above discussion suggests two approaches
to deriving a bulk formula for the OMP $\theta$.
One is to follow (ii) and compute the orbital
magnetization~\cite{thonhauser,niugroup} in an applied electrical
field.  We focus here on (i) instead, working via
$dP/dB$.
The modern theory of polarization starts from the polarization
current $j_P = dP/dt$ under slow deformation of the Bloch
Hamiltonian, and contains, to first order in $d/dt$,
one power of the Berry curvature defined below~\cite{ksv}. 
Using semiclassical wavepacket dynamics,
Xiao {\it et al.}~\cite{xiao} have shown how
to compute the polarization current to second
order and to incorporate slow spatial variations in
the electronic Hamiltonian.
For the case of an orthorhombic 3D crystal
with $M$ occupied bands in which the slow
spatial variation occurs along the $y$ direction in a supercell
of length $l_y$, they obtain
\beq
\ev{\Delta P_x\inh}\!
=  \!\frac{e}{4} \!\int_0^1 \! d\lambda\! \intbz \! \int_0^{l_y} \! \frac{dy}{l_y}
  \,\epsilon_{ijkl} {\rm Tr}[\F_{ij}\F_{kl}]
\label{eq:DPin}
\eeq
for the change in the
supercell-averaged polarization arising from adiabatic currents that
are inhomogeneously induced as a global parameter $\lambda$ evolves
from 0 to 1.  Here indices $ijkl$ run over
$(k_x,k_y,y,\lambda)$,
${\cal F}_{ij} = \partial_i {\cal A}_j - \partial_j
{\cal A}_i -i [{\cal A}_i, {\cal A}_j]$
is the Berry curvature tensor ($\A_\lambda^{\mu\nu}=
i\me{u_\mu}{\del_\lambda}{u_\nu}$),
and the trace and commutator refer to band indices.

Because $\F$ is gauge-covariant,
the integrand in Eq.~(\ref{eq:DPin}) is explicitly gauge-invariant;
it is the non-Abelian second Chern class \cite{nakahara},
so that Eq.~(\ref{eq:DPin}) is path-invariant modulo a quantum
$e/a_zl_y$, where $a_z$ is the lattice constant in the $z$ direction.
Moreover, the $\lambda$ integral can be performed to obtain an
expression in terms of the non-Abelian Chern-Simons 3-form \cite{nakahara}.  Thus,
\beq
\ev{P_x\inh}
=  e \!\intbz \int_0^{l_y}\!\frac{dy}{l_y}\,
\epsilon_{ijk}{\rm Tr}[\A_i\del_j\A_k-\frac{2 i}{3}\A_i\A_j\A_k]
\label{eq:Pin}
\eeq
where $ijk$ now run only over $(k_x,k_y,y)$.  Here the integrand is not
gauge-invariant, but the integral is
gauge-invariant
modulo the quantum $e/a_z l_y$.

We apply this result to study the polarization
\beq \label{PofB}
\ev{P_x\inh}\!=\!\frac{Be^2}{\hbar}\!\intbz  \epsilon_{ijk}
{\rm Tr}[\A_i\del_j\A_k-i\frac{2}{3}\A_i\A_j\A_k]
\eeq
induced by a magnetic field described by the inhomogeneous vector potential
${\bf A}=By\hat{\bf z}$ with $B=h/ea_zl_y$, i.e., a $B$-field along
$\hat{\bf x}$ with one flux quantum threading the supercell.  This
has the effect of taking $k_z\rightarrow k_z+eB y/\hbar$, and this is
the only $y$-dependence in the Hamiltonian,
so that $\ket{\del_yu}=(Be/\hbar)\ket{\del_{k_z} u}$ and
where $ijk$ now run over $(k_x,k_y,k_z)$.
Using Eqs.~(\ref{eq:alpha},\ref{eq:alpha-decomp})
we arrive directly at Eq.~(\ref{theta}).

There is an important geometrical relationship in this (noninteracting) derivation that applies equally well to the many-body case and gives a bulk interpretation of the $2\pi$ ambiguity in $\theta$, whose surface interpretation was in terms of allowed surface IQHE layers.  Polarization in a crystal is defined modulo the ``quantum of polarization'' \cite{ksv} which, for the
flux-threaded supercell of Eq.~(\ref{PofB}),
is $\Delta P_x = e /a_z l_y$.   Since the magnetic field
is $B_x = h/e a_z l_y$, it follows that $\Delta(P_x/B_x)=e^2/h$.
Hence the unit-cell-independent ambiguity of $dP/dB$ results from the relationship in a finite periodic system between the unit-cell-dependent polarization quantum and the quantization of applied flux, and this relationship remains valid in the many-body case.

Before studying the OMP in a specific model, we discuss its symmetry
properties and how to obtain it when Bloch states are unavailable,
as in the many-particle case.
Clearly the combination $\E\cdot\B$ in Eq.~(\ref{axioncoupling})
is odd under $T$ and under inversion $P$ (although it is
even under the combination $PT$).
It is also
odd under any improper rotation, such as a simple mirror reflection.
This implies that $\theta=-\theta$ if the crystal has {\it any}
of the above symmetries.  This would force an aperiodic coupling to vanish,
but since $\theta$ is only well-defined modulo $2\pi$, it actually
only forces $\theta=0$ or $\pi$.  Thus, one can obtain an insulator
with quantized $\theta=\pi$ not only for $T$-invariant systems
(regardless of whether they obey inversion
symmetry), but also for inversion- and mirror-symmetric crystals
regardless of $T$ symmetry~\cite{fu&kane2-2006}.  When none of these
symmetries are
present, one generically has a non-zero (and non-$\pi$)
value of $\theta$, but still retaining the simple scalar
form of Eq.~(\ref{axioncoupling}).

In an interacting system, the OMP should be obtained from the
many-particle wavefunction.  However, modifying Eq.~(\ref{theta})
to the Abelian Chern-Simons integral over the many-body
wavefunction fails~\footnote{Mathematically, obtaining a
nontrivial second Chern or Chern-Simons integral depends on
having a degenerate set of bands somewhere in parameter space,
which is not the case for a gapped many-body wavefunction.},
in important contrast to the case of the
polarization (the integral of $\A$), where such a generalization
works~\cite{ortizmartin}. 
Instead, the OMP can be found using the
change in the many-body polarization due to an
applied magnetic field to compute $dP/dB$, i.e., the many-body
version of the supercell $dP/dB$ calculation.  This fact
is important beyond computing $\theta$ with interactions, as
it defines the topological insulator phase in the many-body
case more simply than before~\cite{leeryu}.  Like the IQHE,
the topological insulator is defined via a response function
($dP/dB$) to a perturbation that, in the limit of a large system
with periodic boundary conditions, is locally weak and hence
does not close the insulating gap.  In the IQHE, this response
function is to a boundary phase (i.e., a flux that does not pass
through the 2D system), while for the topological insulator,
the defining response is to a magnetic flux through the 3D system.

In the remainder of this Letter, we
demonstrate the
above theory via numerical calculations on a tight-binding
Hamiltonian that generates non-zero values of $\theta$, then discuss experimental measurements of $\theta$.  We start with
the model of Fu, Kane, and Mele~\cite{fu&kane&mele-2007} for
a 3D topological insulator on the diamond lattice,
\begin{equation}\label{modelham}
H_{FKM} = \sum_{\langle i j \rangle} t_{ij} c_i^\dagger c_j
 + i \frac{4 \lambda_{SO}}{a^2} \sum_{\langle \langle i j \rangle \rangle}
		 c_i^\dagger \bm{\sigma} \cdot (\bm{d}_{ij}^1 \times \bm{d}_{ij}^2) c_j.
\end{equation}
In the first term, the nearest-neighbor hopping amplitude depends
on the bond direction; we take $t_{ij} = 3t + \delta$ for direction [111]
(in the conventional fcc unit cell of linear size $a$)
and $t_{ij} = t$ for the other three bonds.
The second term describes spin-dependent hopping between pairs
of second neighbors $\langle\langle ij\rangle\rangle$, where
$\bm{d}_{ij}^1$ and $\bm{d}_{ij}^2$ are the connecting first-neighbor
legs and $\bm{\sigma}$ are the Pauli spin matrices.
With $|\delta|<2t$ and $\lambda_{SO}$ sufficiently
large, this model has a
direct band gap of $2|\delta|$.

To break $T$ we add a staggered Zeeman field with opposite
signs on the two fcc sublattices $A$ and $B$,
$\bm{h}\cdot \left( \sum_{i\in A} c_i^\dagger \bm{\sigma} c_i
- \sum_{i\in B} c_i^\dagger \bm{\sigma} c_i \right)$.  We take $|\bm{h}| = m
\sin \beta$ and choose $\bm{h}$ in the [111] direction; setting
$\delta = m \cos \beta$ and varying the
single parameter $\beta$ keeps the gap constant and interpolates
smoothly
between the ordinary ($\beta = 0$) and the topological ($\beta =
\pi$) insulator.

\begin{figure}[!ht]
\includegraphics[width=2.7in]{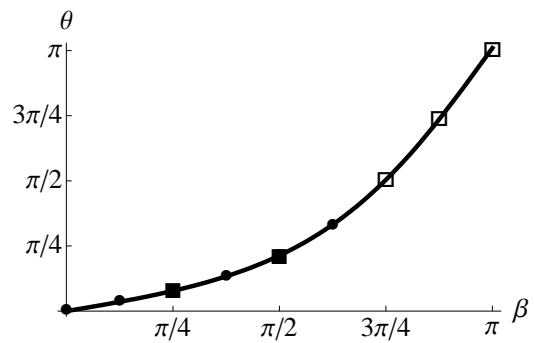}
\caption{ The magnetoelectric polarizability $\theta$
(in units of $e^2/2\pi h$).  The filled squares are computed by
the Chern-Simons form, Eq.~(\ref{theta}).  The open squares are $dP/dB$
from Eq.~(\ref{polarization}).  The points are obtained
by layer-resolved $\sigma_\mathrm{H}$ calculations using
Eq.~(\ref{conductance}).  The curve is obtained from
Eq.~(\ref{secondchernapp}).
\label{polplot} }
\end{figure}

We have calculated the OMP angle $\theta$ using four different methods
with excellent agreement (Fig.~\ref{polplot}).  First, we obtain $\theta$
from Eq.~(\ref{theta}); this requires a smooth gauge for $\A$, which
can be found using now-standard Wannier-based
methods~\cite{MarzariVanderbilt97}.  Results are shown
for $\beta = \pi/4$ and $\beta = \pi/2$
(filled squares).  

Next, we have calculated the polarization~\cite{ksv}
\begin{equation} \label{polarization}
P_i = e \int_{BZ} \! \frac{d^3 k}{(2\pi)^3} \mathrm{Tr}\,\mathcal{A}_i \,.
\end{equation}
resulting from a single magnetic flux quantum in a large supercell.
Varying the supercell size (and thereby $B$)
allows us to approximate $dP/dB$, yielding the open squares in Fig.~\ref{polplot}.
The points in Fig.~\ref{polplot} are from the surface Hall response
in a slab geometry, described below.
Finally, to obtain the curve in Fig.~\ref{polplot}, we
also computed $\theta(\beta)$ from the second Chern
expression~\cite{xiao,qilong}
\beq
\theta = \frac{1}{16\pi} \int_0^\beta d\beta^\prime \int d^3k\,\epsilon_{ijkl} \mathrm{Tr}[{\cal F}_{ij}({\bf k},\beta^\prime) {\cal F}_{kl} ({\bf k},\beta^\prime)]
\label{secondchernapp}
\eeq
(derived above as Eq.~(\ref{eq:DPin})).
Clearly, the various approaches are numerically equivalent.

We now discuss the surface Hall conductivity,
whose fractional part in units of
$e^2/h$ is just $\theta/2\pi$~\cite{wilczekaxion}.
Consider a material with coupling $\theta$ in a slab geometry
that is finite in the ${\bf \hat z}$
direction and surrounded by $\theta=0$ vacuum.  The simplest
interfaces will then lead to $\sigma_{H} = \theta e^2 / (2 \pi h)$
at the top surface
and $-\theta e^2/(2 \pi h)$ at the bottom surface, for a total
$\sigma_{xy}$ of zero.
More generally, arbitrary surface quantum Hall layers change the total integer quantum Hall state,
but not the fractional parts at each surface.

\begin{figure}
 \includegraphics[width=2.7in]{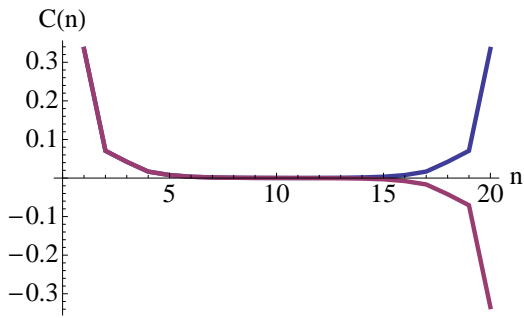}
 \caption{(Color online)
  The layer-resolved Hall conductivity (in units of $e^2/h$) at
  $\beta = \pi$ in a slab of twenty layers, with $m=t/2$ and
  $\lambda_{SO} = t/4$, terminated in $(\bar{1}11)$ planes.
		}  \label{currplot}
\end{figure}

The spatial
contributions to the Hall conductance in the slab
geometry can be resolved as follows.  The unit cell is a supercell containing
some number $N$ of original unit cells in the ${\bf \hat z}$
direction, with translational invariance remaining in the ${\bf
\hat x}$ and ${\bf \hat y}$ directions.  The TKNN integer
for the entire slab
is~\cite{tknn,ass}
\beqn
C =\frac{i}{2\pi} \int\!d^2 k\,{\rm Tr} \left[\mathcal{P} \epsilon_{ij}
\partial_i \mathcal{P} \partial_j \mathcal{P} \right].
\eeqn
Here $i$ and $j$ take the values $k_x$ and $k_y$
and $\mathcal{P}=\sum_\nu |u_\nu\rangle \langle u_\nu|$ is
the projection operator onto the occupied subspace
($\nu$ runs over occupied bands).
To find how different ${\bf
\hat z}$ layers contribute to $C$, define
a projection $\tilde{\mathcal{P}}_n$ onto layer
$n$ within the supercell, and compute
\beq
C(n) = \frac{i}{2\pi} \int\,d^2 k\,{\rm Tr} \left[\mathcal{P} \epsilon_{ij}
(\partial_i \mathcal{P}) \tilde{\mathcal{P}}_n (\partial_j \mathcal{P})
\right]. \label{conductance}
\eeq
The results, presented in Fig.~\ref{currplot}, confirm that the surface
layers have half-integer Hall conductance when $\beta = \pi$ in
(\ref{modelham}) and that the sign on each surface is
switched by local $T$-breaking perturbations (in this example, a uniform
Zeeman coupling in the surface layer).

% This confirms the equivalence of the second Chern definition to
% previous definitions of the 3D topological insulator.

%Showing this equivalence for one model is essentially sufficient to
%show equivalence in general, since both the $\mathbb{Z}_2$ integral
%in $T$-invariant planes and second Chern integral are homotopy
%invariants.

To gain some insight into the microscopic origin of $\theta$ in the 
noninteracting case, using Eq.~(\ref{theta}) we have calculated $\theta$ for a
Hamiltonian that breaks $PT$ (as well as $P$ and $T$) by adding a weak,
uniform (\emph{i.e.}, not staggered) Zeeman coupling.  
For some values of $\beta$ this lifts all degeneracies, enabling us 
to isolate the single-band and interband contributions to $\theta$
and to verify that, because interband contributions are nonzero 
in general, $\theta$ is a property of the whole occupied spectrum 
(unlike polarization, which is a sum of individual band contributions).
A single filled band can have nonzero $\theta$ only if there are 
more than two bands in total~\cite{hopfinsulator}.
% For some values
% of $\beta$ this lifts all degeneracies, enabling us to isolate the
% single-band and interband contributions to $\theta$.  
% A single filled 
% band can have
% nonzero $\theta$ only if there are more than two bands in total~\cite{hopfinsulator}.
% Because interband contributions are nonzero in general, $\theta$ is
% a property of the whole occupied spectrum, unlike polarization,
% which is a sum of individual band contributions.

Experimental detection of $\theta$ is more difficult for a
topological insulator than for a 
generic magnetoelectric insulator
because some
$T$-breaking perturbation is needed to gap the surface state.
Furthermore, a large surface density of states, as in 
Bi$_{0.9}$Sb$_{0.1}$, may complicate the measurement: while
even a weak magnetic field will in principle lead to a gap and
half-integer quantum Hall effect at each surface, the large number
of filled surface Landau levels may make it difficult to isolate
the half-integer part of surface $\sigma_\mathrm{H}$.
In the presence of 
broken discrete symmetries, as in antiferromagnets
or multiferroics, the surface gap
exists naturally and experiments are easier.  For example, the theoretical
methods of this paper could be used to compute the orbital part of
the recently measured $\theta$ in Cr$_2$O$_3$~\cite{expttheta}.

The authors acknowledge useful discussions with A.~Selem and I.~Souza.
The work was supported by the Western Institute of Nanoelectronics
(AME), NSF DMR-0804413 (JEM), and NSF DMR-0549198 (DV).

%\bibliography{../bigbib}
%\end{document}

\end{document}